\documentclass{emulateapj}
\usepackage{natbib}
\usepackage{amsmath}
\usepackage{amssymb} 
\usepackage{epsfig}

\shortauthors{Miniati \& Bell}

\begin{document}
\title{Resistive Magnetic Field Generation at Cosmic Dawn}

\author{Francesco Miniati}
\affil{Physics Department, Wolfgang-Pauli-Strasse 27,
ETH-Z\"urich, CH-8093, Z\"urich, Switzerland; fm@phys.ethz.ch}
\author{A.~R.~Bell}
\affil{Clarendon Laboratory, University of Oxford, Parks Road, Oxford UK OX1 3PU; t.bell1@physics.ox.ac.uk}

\begin{abstract}
  Relativistic charged particles (CR for cosmic-rays) produced by
  Supernova explosion of the first generation of massive stars that
  are responsible for the re-ionization of the universe escape into
  the intergalactic medium, carrying an electric current. Charge
  imbalance and induction give rise to a return current, $\vec j_t$,
  carried by the cold thermal plasma which tends to cancel the CR
  current.  The electric field, $\vec E=\eta \vec j_t$, required to
  draw the collisional return current opposes the outflow of low
  energy cosmic rays and ohmically heats the cold plasma.  Owing to
  inhomogeneities in the resistivity, $\eta(T)$, caused by structure
  in the temperature, $T$, of the intergalactic plasma, the electric
  field possesses a rotational component which sustains Faraday's
  induction. It is found that magnetic field is robustly generated
  throughout intergalactic space at rate of 10$^{-17}-10^{-16}$
  Gauss/Gyr, until the temperature of the intergalactic medium is
  raised by cosmic reionization. The magnetic field may seed the
  subsequent growth of magnetic fields in the intergalactic environment.
\end{abstract}


\section{Introduction}

Massive stars characterized by high emission of ionizing UV photons
are the main contributors to the process of cosmic re-ionization
\citep[e.g.,][]{cife05}. At the end of their life they explode as
Supernovae accelerating, large amounts of CR protons
\citep{krymsky77,axlesk77,bell78a,blos78}.  Since dynamo generation of
magnetic fields operates very quickly within stars ~\citep{Rees06},
stars can magnetize their surroundings through their own magnetized
stellar winds.  Diffusive shock acceleration around Supernovae can,
therefore, take place even without a pre-existing galactic magnetic
field.
Owing to their much higher energy and diffusive mean free path
compared to thermal particles, CRs eventually escape from the parent
galaxy into the intergalactic medium.  Their escape may or may not be
collimated by the galaxy disk (e.g. by the magnetic field in the wind 
around stars or a pre-existing galactic field if it exists). 
In any case, the CR protons carry a
small but important electric current ${\vec j}_c$ which couples them
to the thermal intergalactic medium and induce magnetic fields
generation there.
The phenomena described in this paper depend on the electrical current
carried by the CR.  Most of the current is carried at GeV energies, so
our discussion does not depend upon CR acceleration to TeV or PeV
energies. Also we neglect CR electrons as they are typically two
orders of magnitude less numerous than
protons~\citep{Schlickeiser2002}.

The Larmor radius of a proton of momentum $p$ in a magnetic field $B$
is $r_g=(p/mc)(B/10^{-15}G)^{-1}$kpc.  In our discussion we will
consider fields up to $10^{-16}$G and protons energies up to ~1GeV, so
magnetic field will place very little restriction on intergalactic CR
propagation.  Even a spatial diffusivity unrealistically as small as
the Bohm diffusivity, in which the mean free path is set equal to the
Larmor radius, would allow GeV protons to diffuse a several Mpc in a
$10^{-15}$ Gauss field after 1 Gyr. 
Inter-particle collisions also place no restriction
on propagation since the Coulomb collision time for GeV protons in a
characteristic plasma density $\sim 10^{-4}$cm$^{-3}$ is
$\sim10^3$Gyr.  CR propagation within galaxies is not necessarily as
clear cut since it is possible that the magnetic field, although
largely unknown, may conceivably be much larger than in the
intergalactic medium as result of processes associated with stars and
supernovae.  However, CR appear to diffuse freely around our own
galaxy with present magnetic fields in the $\mu$G range.  Even at
energies in the GeV range, the presumption is that the galactic CR
spectrum is determined by a balance between CR production by SNR and
escape from the galaxy ~\citep{hillas05}.  Moreover, there is
relatively direct evidence that even in the large magnetic field close
to the center of our galaxy CR diffusion, far exceeding Bohm
diffusion, is relatively free ~\citep{dimitrakoudis09}.  Hence we base
our discussion on the assumption that CR produced by early supernovae
escape freely from their host galaxy.  Once GeV protons escape into
the intergalactic medium their propagation is restricted by neither
collisions nor magnetic field.

In this paper we show that the electric currents carried by CRs
protons produced in the first generation of galaxies that are also
responsible for re-ionization of the universe is sufficiently large to
generate significant magnetic fields throughout intergalactic space.
The magnetic field is generated by the curl of the electric field
associated with the return currents induced by the CR particle
propagation. This process operates most efficiently while the
intergalactic medium is cold, and effectively shuts down once
reionization raises the intergalactic gas temperature above $10^4$ K,
i.e. when the universe was about 1 Gyr old~\citep{cife05}.  The
generated magnetic field is typically of order $10^{-17}$G at the end
of reionization.  This magnetic field may eventually be further
enhanced by a turbulent dynamo~\citep{ryuetal08}, although CR currents
may play a further role in magnetic field amplification through
Lorentz's force~\citep{bell04,bell05}.

Magnetic fields are observed in most astrophysical bodies from
planetary scales, to stars and galaxies, and up to the largest
structures in the universe \citep{zeruso83,kulsrudzweibel08}.  
Nearby galaxies
have been known to be magnetized for some time, but recent studies
show that they acquire their fields when the universe was less than
half its present age \citep{bernetetal08,kronbergetal08}.  Since the
growth of large scale magnetic field in a dynamo model is exponential
in time, this poses serious restrictions on the timescale on which
galactic-dynamo must operate.

Evidence for intergalactic magnetic fields is provided by Faraday
rotation measure and diffuse synchrotron radiation which reveal the
existence of $\mu$G strong magnetic fields in the hot plasma of galaxy
clusters~\citep{clkrbo00,cata02}. These probes are considerably less
sensitive to fields in adjacent structures such as small groups and
filaments of galaxies. Nevertheless, these fields are 
expected to be there and large efforts are being made in an attempt
to measure them. In addition, intergalactic
magnetic fields affect the propagation of ultra high energy cosmic ray
particles, introducing potentially measurable effects on their energy
spectrum, arrival direction and
composition~\citep{sme04,dolag05,hooperandtaylor10}.  Finally, the
presence of magnetic fields in voids can affect the observed spectrum
of extragalactic TeV $\gamma$-ray sources.  Multi-TeV photons are
absorbed by the diffuse extragalactic background light and converted
into $e^\pm$ pairs which emit secondary cascade multi-GeV
$\gamma$-rays by inverse Compton on the cosmic microwave background.
The observed flux of pair produced GeV photons can be suppressed if
the pairs are deflected from the line of sight by a magnetic field
within an energy loss distance~\citep{ahcovo94,neronovsemikov09}.
Based on these ideas, in very recent studies\footnote{these results
  appeared after the first submission of this paper} the $\gamma$-ray
spectra of blazars were used to set lower limits to the value of
magnetic fields in voids in the range $5\times
10^{-15}-5\times10^{-17}$G~\citep{neronovandvovk10,tavecchio10}.

The presence of magnetic fields in astrophysical plasma requires a
mechanism for their generation because, given the high mobility of the
charges, sustained electrostatic fields are non-trivial to set up.
However, even a weak seed may suffice because magnetic field
amplification can occur at the expense of the plasma motions through
the induced electric field, $\vec E=-\vec v\times \vec B$.
Specific lower limits on the required seed strength depend on the
astrophysical systems.

There already exist various scenarios for the generation of magnetic
fields each with its strengths and weaknesses, among others Weibel's
instability at shocks~\citep{msk06,schsh03,medvedev07}; battery
effects, e.g. due to Compton drag \citep{harrison70,ichikietal06} or
Biermann's mechanism either at cosmic shocks or ionization fronts
\citep{subramanian94,kcor97,gnfezw00}; galactic winds
~\citep{bertone06,donnertetal2009}, processes in the early
universe~\citep{widrow02}.
In fact, the magnetization of cosmic space is complex and various
processes are likely to contribute to it, though to a different extent
in different environments.  One of the strengths of the model we
describe below is that it generates relatively strong macroscopic
fields throughout cosmic space, while having a simple description
based on well understood physical processes.

The remainder of this paper is organized as follows: In
Section~\ref{resistive:se} we introduce the resistive mechanism for
the generation of magnetic field.  In Section \ref{simul:se} we
describe a cosmological simulation to determine some of the parameters
describing the intergalactic medium that are required to quantify the
generation of magnetic field.  In Sections \ref{galsol:se} and
\ref{fixedjsol:se} we discuss the generation of magnetic field around
individual galaxies using a numerical and slightly simplified
analytical solutions, respectively. The solution is extended to the
case of the intergalactic space in \ref{igmsol:se}. Finally, we
briefly summarize the main findings of this paper in~\ref{summary:se}.

In the following we use SI units and, unless explicitly stated,
lengths are expressed in physical units, not comoving units.

\section{Resistive mechanism} \label{resistive:se}

Though, as discussed in the previous section, CR propagation in
intergalactic space is unaffected by Coulomb collision and magnetic
fields deflections, CRs are not completely decoupled from
intergalactic plasma.  This is due to the small current, ${\vec j}_c$,
carried by the CR flux, which is assumed mainly to consist of
protons. In fact, to maintain quasi-neutrality the CR current must be
balanced by a return current carried by the thermal plasma ${\vec
  j}_t$.  While the CR are collisionless with very long mean free
paths, the thermal particles, due to their low energy, have mean free
paths shorter than scale lengths of interest here.  For example, at
redshift $z\simeq10$, when the intergalactic plasma had a temperature
$\sim 1$K and density $\sim 10^{-4}$cm$^{-3}$, thermal electrons had a
Coulomb mean free path of only $10^5 \psi^{-1}$m where $\psi$ is the
fractional ionization.
Consequently an electric field in the rest frame of the plasma (indicated
by a $\prime$) is required to draw the return current, namely
\begin{equation} \label{ohm:eq}
\vec{E}^\prime = \eta \vec j_t,
\end{equation}
where $\eta$ is the plasma resistivity.
Since Coulomb collisions dominate over charge-neutral
collisions the resistivity takes approximately the Spitzer value,
$\eta =65 T^{-3/2}${\rm log}$ \Lambda$ $\Omega $m, where $T$ is the
temperature of the ambient thermal plasma in K, and log$\Lambda\simeq 20$ is
the Coulomb logarithm. The Spitzer resistivity is independent of
electron number density and consequently independent of the degree of
ionization.

The electric field transfers energy from the CR flux to the thermal
plasma through ohmic heating, described by the equation
\begin{equation} \label{ohmic:eq}
\frac{3}{2}nk_B\frac{dT}{dt}=\eta j_c^2
\end{equation}
where $k_B$ is the Boltzmann constant.  Since the charge-neutron
collision time is much smaller than the expansion time of the
universe, in Eq. (\ref{ohmic:eq}) $n$ is the total number particles in
the plasma, to include heating of neutral as well as ionized
particles.  As a result of ohmic heating the temperature of the
initially cold thermal plasma is raised.

The electric field also opposes the current carried by the CRs.  In
particular, CRs escape the parent galaxy to a distance $R$ only if
their kinetic energy at source exceeds the potential at $R$, $\phi =
\int^R_0 | E | dR$.  This can determine the minimum energy of CR reaching
$R$, giving
\begin{eqnarray} \label{pmin:eq}
p_{min}=\frac{e\phi(R)}{c} \left(1+2\frac{mc^2}{e\phi(R)}\right)^{1/2}
\end{eqnarray}
As will be seen in figure 3, the electric potential is not sufficient
to limit the escape of mildly relativistic protons from the galaxy for
the cases considered here, largely because we assume that gas within
100kpc of a galaxy is hot (see below), and therefore has a low resistivity which
results in a low electric field even though the CR current is
relatively large close to the galaxy.

In a uniform medium, the electric field drawing the return current $\vec
j_t$ cannot have a curl and therefore there is no generation of
magnetic field. If the medium is non-uniform on a given scale, the
forward and return currents can become separated on that
scale producing a current loop which in turn supports a magnetic field.
However, the separation of forward and return currents, and the
corresponding magnetic field, is opposed by inductive effects, and the
actual growth of magnetic field must be determined by self-consistent
solution of the Maxwell equations
\begin{eqnarray} \label{ampere:eq}
\nabla \times {\vec B}&=&\mu_0 (\vec j_c+\vec j_t), \\ \label{faraday:eq}
\frac{\partial {\vec B}}{\partial t} &=& -\nabla \times {\vec E}.
\end{eqnarray}
The displacement current can be neglected because timescales are much
longer than the light transit time and the system evolves through a
series of quasi-neutral steady states.  Using Amp\`ere's law
(\ref{ampere:eq}) to eliminate the thermal current we find that the
electric field (\ref{ohm:eq}) in the frame in which the the plasma
moves at velocity $\vec v$, is $\vec{E} = -\vec
v\times\vec{B}+(\eta/\mu_0)\vec\nabla\times\vec{B}- \eta \vec j_c$.
The curl of this electric field then produces growth of magnetic field
according to Faraday's law\footnote{Adiabatic losses due to cosmic
  expansion are negligible during the $\sim$ 1 Gyr in which the
  resistive process is efficient.}
\begin{equation}
\frac{\partial\vec B}{\partial t}= \vec\nabla\times(\vec v\times\vec B)
-\vec\nabla\times \left(\frac{\eta}{\mu_0}\vec\nabla\times\vec{B} \right )
+\vec\nabla\times(\eta\vec j_c ).
\end{equation}

The first term on the right hand side transports the frozen-in field
with the plasma and can stretch and amplify an already existing
magnetic field. The second term represents resistive diffusion
and can easily be verified to be insignificant for kpc distances and
Gyr timescales.
Crucially, in contrast, the final {\it resistive}
term of the equation produces magnetic field in a previously
unmagnetized plasma. The magnetic
field grows wherever the resistivity varies perpendicularly to the CR
current.  A larger electric field is needed to draw the return current
where the resistivity is higher, so the electric field has a curl and
magnetic field grows.  These equations are well known in
laser-produced plasmas where the current is carried by energetic
laser-produced electrons in place of Supernovae-produced CRs 
protons~\citep{BellKingham2003}.

Temperature inhomogeneities, resulting in variations in conductivity,
are naturally present throughout intergalactic plasma due to the
growth of cosmological structure.
In addition, because the volume heat capacity is proportional to
density, where ohmic heating is significant, inhomogeneous temperature
enhancements will arise from density inhomogeneities.  On the other
hand, the CR current is expected to be approximately uniform because
as discussed previously sections, CR are relatively undeflected by
collisions or magnetic fields in the intergalactic medium.  Thus the
magnetic field may be generated by the resistive term at a rate 
\begin{equation}\label{bdot:eq}
\dot B=\vert \vec j_c\times\vec\nabla\eta\vert \simeq 
\frac{j_{c}\eta}{L_{T}},
\end{equation}
where, $L_T\equiv T/|\nabla T|$, is the characteristic
temperature scale.

It is clear from the above expression that 
the resistive process depends sensitively on the plasma temperature
through $\eta$.  In particular, it operates efficiently while the
intergalactic medium is cold, and it effectively shuts down once
reionization raises the intergalactic gas temperature above $10^4$
K. For a universe that is reionized at redshift $z\ge
6$~\citep{cife05}, the available time is of order a Gyr.  Note that
close to the star-forming galaxies the gas temperature is $\ge 10^4$K due
to both photoheating from the ionizing flux escaping the galaxies and
the ohmic heating.  For the above choice
of the reionization epoch, the typical size of HII regions at $z\simeq
10$ is about 100 kpc \citep[or 1 comoving Mpc,][]{zahnetal07}.

In order to quantify the importance of the resistive process in the
following we first use a numerical simulation of structure formation
to compute the scale of the temperature variations in the
intergalactic medium. We then estimate the generation of magnetic
field around individual star-forming galaxies at the epoch of
re-ionization, including self-consistently the effect of the
induced electric fields on the escaping CR particles and ohmic
heating. Finally, we estimate the distribution of magnetic fields
produced in intergalactic space.

\section{Cosmological simulation}\label{simul:se}

We extract characteristic scale-length $L_T$ and the characteristic
range of values taken by $L_T$ from a cosmological simulation of
structure formation which includes hydrodynamics, the relevant
thermodynamic processes for the diffuse baryonic gas, dark matter and
gravity. The simulation does not include the cosmic ray effects
discussed in this paper.  The simulation uses a directionally un-split
higher order Godunov's method for the hydrodynamics, a time centered
modified symplectic scheme for the collisionless dark matter and we
solve Poisson's equation with a second order accurate discretization
\citep{mico07b}.  Atomic cooling, with rates as given
in~\cite{huignedin97}, is included according to the
algorithm proposed in~\citep{mico07a,miniati10}, although prior to
reionization adiabatic expansion of the universe is the most relevant
cooling process.  We adopt a flat $\Lambda$CDM universe with the
following parameters normalized to the critical value for closure: 
total mass density, $\Omega_m=0.2792$, baryonic mass density,
$\Omega_b=0.0462$, vacuum energy density, $\Omega_\Lambda= 1-
\Omega_m= 0.7208$~\citep{komatsuetal09}. In addition,
the normalized Hubble constant is
$h\equiv H_0/100$ km s$^{-1}$ Mpc$^{-1}$ = 0.701, the
spectral index of primordial perturbation, $n_s=0.96$, and the
rms linear density fluctuation within a
sphere with a comoving radius of 8 $h^{-1}$ Mpc,
$\sigma_8=0.817$~\citep{komatsuetal09}.  We generate the initial
conditions with the {\rm grafic2} package~\citep{bertschinger01}.  We
use a computational box of comoving size $L=1h^{-1}$ Mpc, discretized
with 512$^3$ comoving computational cells, providing a nominal spatial
resolution of 2$h^{-1}$ comoving kpc for the field components, and
512$^3$ particles with mass $4.8\times 10^2h^{-1}$ M$_\odot$ for the
collisionless dark matter component.  At redshift $z=6$ the box size
in physical units is $\simeq$ 143 $h^{-1}$kpc and, likewise, the nominal
spatial resolution is $\simeq$ 285 $h^{-1}$pc.  
Thus, structures with spatial scales
between 1-10 kpc should be adequately resolved for our purposes.

\begin{figure}[t] 
\plotone{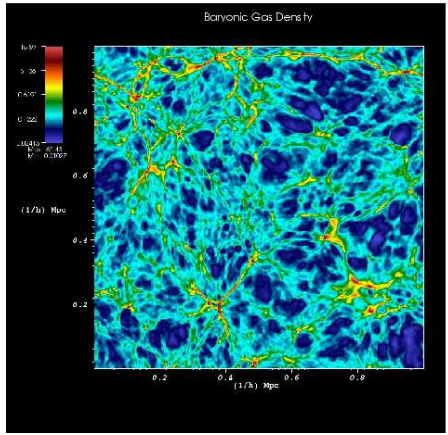}
   \caption{
   Baryonic gas density distribution.}
   \label{rho:fig}
\end{figure}

\begin{figure}[t] 
\plotone{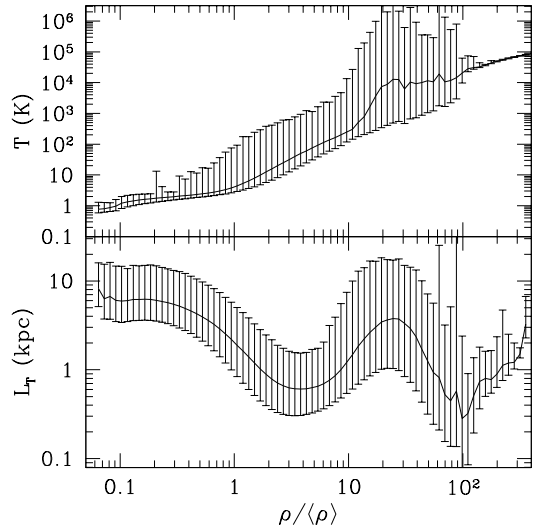}
   \caption{
     Temperature (top) and its characteristic length scale (bottom) as
     a function gas density in units of its average value. The
     vertical bars represent the asymmetric root-mean-squared
     fluctuations about the mean.}
   \label{tmp:fig}
\end{figure}

Figure~\ref{rho:fig} shows the distribution of the baryonic gas
density at $z\simeq10$ on a two dimensional slice across the
simulation box.  One can recognize a few high density collapsed
structures where gas is rapidly cooling. These are the sites where
stars and galaxies form and CR are eventually produced. However, most
of the gas ($\sim$ 97\% by mass) is still in the diffuse phase, with a
density within a factor a few of the mean value.  Figure~\ref{tmp:fig}
shows the occurrence of spatial scale-lengths in the range 1-10 kpc in the
low density, low temperature gas which occupies most of the
intergalactic space.

Note that the residual fraction of free electrons in the intergalactic
medium after recombination keeps the gas temperature $T$
locked to the CMB temperature, $T_{CMB}=2.725(1+z)$ K, until
redshift $1+z=142(\Omega_b h^2/0.024)^{2/5}\simeq 140$ \citep{peebles93}.
After that the temperature of the diffuse intergalactic medium
drops due to adiabatic expansion following the $\gamma-$law for
a monoatomic gas with $\gamma=5/3$.

\section{Solution around individual galaxies}\label{galsol:se}

The growth of magnetic field around an individual galaxy is described
by Eq. (\ref{bdot:eq}). To solve this equation in the following we use
the temperature scale-length derived from the simulation in the
previous section as a reference value for $L_T$.  In addition, we will
determine the CR current produced by a galaxy of a given luminosity.
The current determines the electric field which sets the minimum
momentum of the escaping CRs, Eq. (\ref{pmin:eq}), from which the
current itself depends. In addition, the electric field is effectively
related nonlinearly to the current because the resistivity is affected
by ohmic heating, Eq.~(\ref{ohmic:eq}).  So all these equations need
to be solved self-consistently.

The luminosity of a typical bright galaxy at redshift $z\ge6$, is
$L_*\simeq 5.2\times 10^{21}\mbox{\,W\,Hz}^{-1}$, corresponding to a
star formation rate $\dot
M_*=6.5\,M_\odot\,$yr$^{-1}$~\citep{bouwens07}.  If we conservatively
assume a Salpeter initial mass function, the energy released by SN
explosions per unit mass of formed stars is, $E_{SN}\simeq 5.4\times
10^{42}\mbox{\,J\,}M^{-1}_\odot$. A fraction, $\epsilon_{c}\simeq
30\%$, of this energy is typically converted into CR particles,
implying a CR-luminosity, $L_{c}\simeq 3.2\times 10^{35}
\mbox{\,W}\left(\epsilon_{c}/0.3\right)\left(L/L_*\right)$, for a
galaxy of luminosity $L$.  The corresponding CR energy flux at a
distance $R$ from the galaxy is $Q_c \simeq 2.7\times 10^{-5}
\left(\epsilon_{c}/0.3\right)\left(L/L_*\right) \left(R/{\rm
    kpc}\right)^{-2} {\rm Wm}^{-2}$, where we have idealized the
galaxy as an isotropic point source of CR. If the CRs are collimated
within a solid angle $\Omega_c$ then the CR current intensity is
higher by a factor $4\pi/\Omega_c$, although this does not change the
qualitative picture.  The CR momentum distribution can be expected to
be $p^{-2.3}$ as typically observed for CR production by shocks in the
Galaxy.  With this power-law, most of the energy resides in mildly
relativistic protons. The electric current is predominantly carried by
the lowest momentum CR. In the following analysis, the electric field
is allowed to inhibit CR propagation, although we find that inhibition
is negligible for our conditions.  If only CR with momentum greater
than $p_{min}$ reach a radius $R$, the current carried by CR at that
radius is approximately given by $j_c=5\times
10^{-10}Q_c(p_{min}/m_pc)^{-0.3}/(p_{min}/m_pc+1) $ Amp m$^{-2}$ where
$m_p$ is the proton mass.  Taking the efficiency of CR production to
be fixed at $\epsilon _c=0.3$
\begin{eqnarray}\nonumber
j_c\simeq  5.3\times 10^{-14}
\left(\frac{L}{L_*}\right)
\left(\frac{R}{\rm kpc}\right)^{-2}
\left(\frac{p_{min}}{m_pc}\right)^{-0.3}\\
\left(1+\frac{p_{min}}{m_pc}\right)^{-1}
{\rm Amp\ m}^{-2} \label{jc:eq}
\end{eqnarray}
This CR current must be balanced by an equal but opposite return
current drawn by an electric field $E=\eta j_c$, giving
\begin{eqnarray} \nonumber
E\simeq 6.9\times 10^{-11}
\left(\frac{L}{L_*}\right)
\left(\frac{R}{\rm kpc}\right)^{-2}
\left(\frac{p_{min}}{m_pc}\right)^{-0.3} \\
\left(1+\frac{p_{min}}{m_pc}\right)^{-1}
\left(\frac{T}{\rm K}\right)^{-3/2}
{\rm V\ m}^{-1}. \label{e:eq}
\end{eqnarray}

So we can now solve numerically the coupled system of equations
(\ref{ohmic:eq}), (\ref{pmin:eq}),~(\ref{jc:eq}),~(\ref{e:eq}), to
determine the propagation of CR through the intergalactic medium
surrounding a star forming galaxy, the associated electric field and
the ensuing generation of magnetic field.  Our 'standard' calculation
is as follows.  We treat the galaxy as a spherically symmetric source
of CR which is constant in time.  We set an inner radial boundary to
the computational grid at $r_{min}=10$kpc which corresponds to the
notional radius of the galaxy.  We assume that all CR produced by the
galaxy escape through this boundary for the reasons given in Section
1.  As will become clear below, changing $r_{min}$ has very little
effect on the results for the intergalactic medium on Mpc scales. The
CR flux is specified at the inner boundary as a power law ($\propto
p^{-2.3}$) with a minimum momentum of $0.1m_pc$ which corresponds to a
proton velocity a few times the shock velocity of an expanding young
supernova remnant.  We set the initial intergalactic temperature to
$2\times 10^4$K within a radius of 100kpc to represent heating due to
ionization by the galaxy.  Outside this radius we set the initial
temperature to 1 K, with the temperature changing between the two
regions over a distance of 10kpc.  We set the intergalactic density to
$10^{-4}$cm$^{-3}$.

\begin{figure}[t] 
\plotone{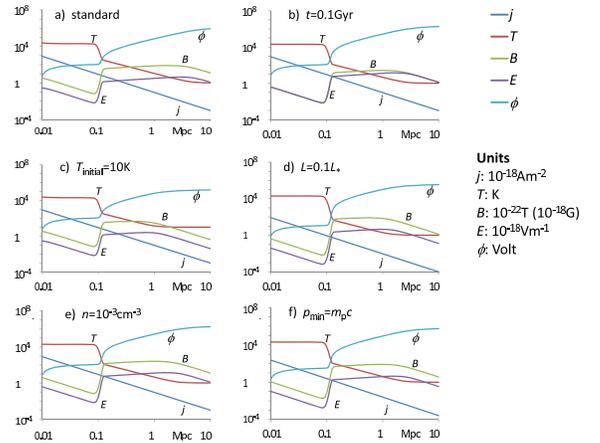}
   \caption{Spatial profiles at radii between 10kpc and
10Mpc.  a) standard calculation, b)-f) as for the standard calculation with one parameter changed as indicated.
The units are
CR current density $j$ in $10^{-18}$Am$^{-2}$, 
temperature $T$ in degrees K,
magnetic field $B$ in $10^{-22}$T ($10^{-18}$G),
electric field $E$ in $10^{-18}$ Vm$^{-1}$,
electric potential $\phi$ in Volt. }
\end{figure}

\subsection{CR propagation and magnetic field generation}
Figure 3a shows the results for our standard calculation, as defined
above, for CR propagation and plasma heating as a function of radius
for a galaxy with typical luminosity $L=L_*$ at time $t=$ 1 Gyr.  The
horizontal scale is logarithmic and stretches from the edge of the
galaxy at 10kpc to a distance of 10Mpc which is characteristic of the
distance between bright galaxies at this epoch.  Ohmic heating is
effective out to a radius of $\sim 3$Mpc where the electric and
magnetic fields reach their maximum values.  The magnetic field is
less within this radius because ohmic heating reduces the resistivity
and the growth of magnetic field.  Within 100kpc the magnetic field is
much smaller because the temperature is initialized to a much larger
value to represent ionization by the galaxy.  The magnetic field is
largest at $\sim 3$Mpc, but the maximum is broad, and fields above
$10^{-17}$G extend from 100kpc to 10Mpc.  Figure 3b displays the
profiles at the earlier time of 0.1Gyr for the same parameters.  The
fields are slightly reduced but still reach $2.5\times 10^{-17}$G,
once again with a broad maximum.  This shows that most of the field
generation occurs quickly before ohmic heating reduces the
resistivity.  Fields in the range $10^{-17}-10^{-16}$G are produced
even if suitable conditions for growth apply for much less than 1Gyr.

We take figure 3a as the standard calculation and then vary individual
parameters in figures 3c-3f.  Fields of a similar magnitude are
produced if the initial temperature is raised to 10K (figure 3c), the
CR luminosity of the galaxy is reduced tenfold (figure 3d), the
density is increased to $10^{-3}$cm$^{-3}$ (figure 3e), or the minimum
CR energy is increased to $p_{min}={\rm m_p c}$ (figure 3f).  Clearly
the production of fields in the range $10^{-17}-10^{-16}$G is robust.
The results indicate that fields as large as $\sim 10^{-20}$Tesla
($10^{-16}$Gauss) are produced in a few to several $\sim$Mpc$^3$ volume
surrounding galaxies if the temperature or density vary on kpc scales.

The plots of the electric potential $\phi$ in figure 3 show that the
electric field does not inhibit the escape of CR from the galaxy.

\section{Analytic solution for a fixed current}\label{fixedjsol:se}

For the cases considered above, the electric potential does not
inhibit CR transport. If the galaxy can be assumed to produce CR at a
constant rate, the CR current is constant at any point outside the
galaxy.  The evolution of the temperature and magnetic field at that
point is then determined by the equations

\begin{equation}
\frac {dB} {dt}= \frac{\eta j_c}{L_T} 
\ \ \ {\rm ;}\ \ \   
\frac {dT} {dt}= \frac {2\eta j_c^2} {3nk_B}
\end{equation}

The solution is 
\begin{equation}
B=\frac{3}{2} B_1 \frac{j_1}{j_c}\left( \left (1+\frac{5}{3} \frac{j_c^2}{j_1^2} \right ) ^{2/5}-1 \right )
\end{equation}

\noindent
where $T_1$ is the initial temperature in Kelvin, $B_1=(\eta_1
nk_BT_1t/L_T^2)^{1/2}$, $j_1=(nk_BT_1/\eta _1 t)^{1/2}$, and $\eta_1$
is the resistivity at temperature $T_1$. The magnetic field is largest
when $j_{c}=5.15j_1$.  When $j_{c}$ exceeds $5.15j_1$, ohmic heating
increases the temperature, and reduces the resistivity, which in turn
reduces the growth of magnetic field.  Ohmic heating is marginally
important when $j_{c}=5.15j_1$.  At this CR current the magnetic field
achieves a maximum value, $B_{\rm max}=1.046B_1$.  Expressed in more
meaningful terms, after a time $t$ in Gyr, the maximum field is
\begin{equation}
B_{\rm max}
=8.2\times 10^{-17}
\left(\frac{L_T}{\mbox{kpc}}\right)^{-1}
\left(\frac{T_1}{\mbox{K}}\right)^{-1/4}
\end{equation}
$$
\times
\left(\frac{n}{{10^{-4}{\rm cm}^{-3}}}\right)^{1/2}
\left(\frac{t}{{\rm Gyr}}\right)^{1/2}
{\rm G}.
$$
The maximum field occurs where the CR current ($j_{c}=5.15j_1$) is
\begin{equation}
j_{\rm max}
=3\times 10^{-20}
\left(\frac{n}{{10^{-4}{\rm cm}^{-3}}}\right)^{1/2}
\left(\frac{T_1}{\mbox{K}}\right)^{5/4}
\end{equation}
$$
\times
\left(\frac{t}{{\rm Gyr}}\right)^{-1/2}
{\rm Am}^{-2}\ ,
$$
and the maximum field occurs at a distance $R_{\rm max}$ from a galaxy with luminosity $L$, where
\begin{equation}
R_{\rm max}=1.9
\left(\frac{n}{{10^{-4}{\rm cm}^{-3}}}\right)^{-1/4}
\left(\frac{T_1}{\mbox{K}}\right)^{-5/8}
\left(\frac{L}{L_*}\right)^{1/2}
\end{equation}
$$
\times
\left(\frac{p_{\rm min}}{0.1m_pc}\right)^{-0.15}
\left(1+\frac{p_{\rm min}}{0.1m_pc}\right)^{-1/2}
\left(\frac{t}{{\rm Gyr}}\right)^{1/4}
{\rm Mpc}
$$
in a medium with a proton density $n$.  With the proviso that $R_{\rm
  max}$ lies in a realistic range, $B_{\rm max}$ is independent of
the emissivity of the CR source, the CR momentum $p_{\rm min}$ and the
fraction of CR escaping the galaxy.  $B_{\rm max}$ is proportional to
the inverse of the temperature scale-length, but all other dependences
are relatively weak. The weak dependence on the initial temperature
implies that significant magnetic field is generated even if the
initial temperature is greater than 1K.  Hence, the characteristic
magnetic field is robustly of the order of $10^{-17}$ to $10^{-16}$G
when the effect of Ohmic heating is included in the calculation.  The
maximum magnetic fields and their spatial locations in figure 3 obey
these equations.  The weak dependencies on various parameters results in
the broad extent of the maxima in figure 3.

\section{Solution for the intergalactic medium}\label{igmsol:se}

The number density of star forming galaxies at redshift $z\ge6$,
with luminosity, $L$, and per luminosity interval, $dL/L_*$,
is well described by a Schechter function
\begin{equation}\label{schechter::eq}
\Phi(L) = \Phi_*\left(\frac{L}{L_*}\right)^{-\alpha} \mbox{\rm e}^{-L/L_*}
\end{equation}
with the following parameters: $L_*\simeq 5.2\times
10^{28}\mbox{\,erg\,s}^{-1}\mbox{\,Hz}^{-1}$, $\Phi_*\simeq
10^{-3}\mbox{Mpc}^{-3},$ $\alpha\simeq 1.77$
\citep{bouwens07,oesch09,bouwens10}.  $L_*$ is the typical luminosity
of a bright galaxy and $\Phi_*$ roughly corresponds to their number
density. $\alpha$ is the slope of the distribution at the faint
luminosity end.

The electric current at any given point in intergalactic space is
contributed by galaxies within the current horizon, $R_{jh}$.  As long
as the diffusion coefficient remains much larger than Bohm's value, as
observed in the much more turbulent interstellar medium of the
Galaxy~\citep{droka09}, even for $B\sim 10^{-16}$G, $R_{jh}$ is of
order a few tens of Mpc, i.e.  larger than the average distance
between bright galaxies.  Magnetic field generation is thus expected
to be dominated by the nearest luminous galaxy.  This is the case even
if the currents are beamed with an opening angle $\theta\sim
0.5$ rad.  In this case, the number of CR current sources visible
from any given point is,
\begin{equation}
N_c=\Phi_*\Gamma(2-\alpha)R_{jh}^3(\theta^2/4\pi)\sim 1,
\end{equation}
for $R_{jh}\ge$ a few $\times$ 10 Mpc. Therefore, each point in space
is exposed to the CR current from about one galaxy.  Note that because
the faint-end slope of the above Schechter's function is steep, i.e.
$\alpha\le 2$, the CR
output per luminosity log-interval,
$\epsilon_cL_*(L/L_*)^2\Phi(L)\simeq \epsilon_cL_*(L/L_*)^{2-\alpha}\Phi_*$, is
approximately constant.  However, the average distance between faint
galaxies of luminosity $L$ is
\begin{equation}
\langle d_L\rangle=\left[L\Phi(L)\right]^{-1/3}\propto L^{(\alpha-1)/3},
\end{equation}
so that their CR current at this distance scales as 
\begin{equation}
j_c(L,d_L)\propto\epsilon_c L/\langle d_L\rangle^2\propto L^{1/2},
\end{equation}
i.e.  fainter galaxies are slightly less
efficient at magnetizing their surroundings.

\begin{figure} [t] 
\epsscale{1}\plotone{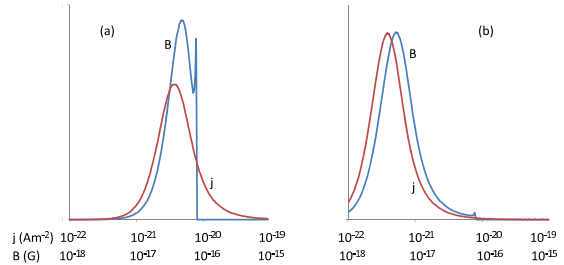}
\caption{a) Distribution functions of current and magnetic field for
  galaxies randomly distributed in space according to the Schechter
  function.  Apart from the luminosity the Monte Carlo calculation
  adopts the standard parameters used in figure 3a. b) distribution
  functions when the CR luminosity of all galaxies is reduced
  tenfold.}
\end{figure}

We now make a more quantitative estimate through a Monte-Carlo
simulation in which galaxies are distributed randomly in space
according to the Schechter function as given in equation
(\ref{schechter::eq}). The
currents from each galaxy are added vectorily and the calculation
repeated many times to build-up a distribution function for the CR
currents found at any random point in space. The
magnetic field is then calculated using equation 11. Figure 4a gives
the distribution functions for the current and the magnetic field for
the standard parameters assumed in figure 3a.  The distribution
functions are defined as the number per logarithmic interval in $j$ or
$B$.  The magnetic field calculation assumes that the initial
temperature is 1K everywhere.  It neglects heating during ionization,
which was included in figure 3 within a radius of 100kpc, but this
neglect has little effect on the distribution function, firstly
because the ionized volume is relatively small compared with the
characteristic distance of 10Mpc between luminous galaxies, and
secondly because, even starting from a low temperature, ohmic heating reduces the resistivity, and therefore
the magnetic field, in regions close to galaxies.  For fixed
parameters such as the temperature scale-length $L_T$ and the plasma
density $n$, the magnetic field cannot exceed a maximum value as given
in equation 12.  This produces an abrupt cut-off to the distribution
function.  The spike in the distribution function at the cut-off in figure 4a
reflects the broad spatial extent of the maxima in figure 3a. The
magnetic field is close to the maximum value over a wide range of
distances from individual galaxies.  In reality, as shown in figure 2,
the temperature scale-length $L_T$ takes a range of local values, and
this will smear the cut-off.  The effect of decreasing (increasing)
the temperature scale-length $L_T$ by a factor $\lambda$
would be to move the distribution function for $B$ a corresponding 
factor  $\lambda$ to the right (left) in figure 4a,
while leaving the distribution function for $j$ unchanged.

In figure 3 it can be seen that the value of the maximum magnetic field
is insensitive to changes in various parameters.  However the position
of the maximum moves in radius.  For example, the maximum magnetic field
occurs at 1Mpc when the CR luminosity of all galaxies is reduced by a
factor ten (see figure 3d where $L=0.1L_*$).  The effect of this on the
distribution function for $B$ can be seen in figure 4b. When the
luminosity is reduced, there are still regions of space in which the
magnetic field reaches the maximum value of $8\times 10^{-17}$G, but
the field is characteristically $10^{-17}$G throughout most of the volume
between galaxies.  Similarly, the effect on the distribution function
of changing other parameters can be deduced from figure 3 in
conjunction with equation 12.  Equation 12 gives a good estimate of
the maximum magnetic field, while figure 3 indicates the volume of space
that is filled by fields close to the maximum value.

\section{Summary}\label{summary:se}

In conclusion, we have shown that cosmic rays produced by supernovae
at the epoch of reionization when the first stars and galaxies form
can be expected to have a substantial impact on the intergalactic
medium out to distances of a several Mpc from the galaxy.  Through the
return electric current carried by the thermal plasma, the medium is
heated ohmically and magnetic field is generated in the range
$10^{-17}$ to $10^{-16}$Gauss.  Because of self-compensation through
ohmic heating, the maximum magnetic fields are relatively insensitive to
parameters such as the galaxy luminosity, the efficiency of CR
production, and the ambient intergalactic temperature and density.
The magnitude of the magnetic field is however sensitive to the
scale-length $L_T$ of temperature variations in the intergalactic
medium.  Cosmological simulations support an assumption of
scale-lengths in the range 1-10kpc.  Self-consistent cosmological
simulations including cosmic ray effects are a necessary next step in
developing this theory, but our argument extends the possibility that
resistive magnetic field generation, driven by cosmic ray streaming,
may generate seed magnetic fields which might subsequently be
turbulently amplified to form the intergalactic magnetic field at the
present epoch.

\acknowledgments
FM acknowledges several useful discussions with Dr. P. Oesch.

\bibliographystyle{apj}
\bibliography{books,papigm,codes,papers,proceed}

\end{document}